\documentclass[aps,prb,citeautoscript,twocolumn,footinbib,notitlepage]{revtex4-1}
\usepackage[a4paper,top=1.5cm,bottom=2.0cm,left=2.0cm,right=2.0cm]{geometry}
\usepackage{amsmath}
\usepackage{amssymb}
\usepackage{bbm}
\usepackage{bm}
\usepackage{color}
\usepackage{graphicx}
\DeclareGraphicsExtensions{.eps, .pdf, .jpg, .png}
\graphicspath{{./}}
\usepackage[pdftex,colorlinks=false,linkcolor=blue]{hyperref}
\usepackage{float}
\usepackage{multimedia}
\usepackage{mathptmx}
\newcommand{\tinyonehalf}{\frac{\mbox{\tiny 1}}{\mbox{\tiny 2}}}
\def\cL{{\mathcal L}}
\def\cM{{\mathcal M}}
\def\cC{{\mathcal C}}
\def\cP{{\mathcal P}}
\newcommand{\He}{$^3$He}
\newcommand{\Heb}{$^3$He-B}
\newcommand{\Hea}{$^3$He-A}
\newcommand{\Hefour}{$^4$He}
\newcommand{\sro}{Sr$_{2}$RuO$_{4}$}
\def\vA{{\bf A}}
\def\vB{{\bf B}}
\def\vd{{\bf d}}
\def\ve{{\bf e}}
\def\vx{{\bf x}}
\def\vy{{\bf y}}
\def\vz{{\bf z}}
\def\vl{{\bf l}}
\def\vm{{\bf m}}
\def\vn{{\bf n}}
\def\vp{{\bf p}}
\def\vr{{\bf r}}
\newcommand{\kb}{k_{\text{B}}}
\newcommand{\grad}{\pmb\nabla}
\newcommand{\vell}{\pmb{\ell}}
\def\ns{\negthinspace}

\def\G{{\mathsf G}}
\def\point#1#2{{\sf #1}_{\mbox{\tiny #2}}}
\def\orbital{{\sf SO(3)_{\text{L}}}}
\def\spin{{\sf SO(3)_{\text{S}}}}
\def\gauge{{\sf U(1)_{\text{N}}}} 
\def\parity{{\sf P}} 
\def\time{{\sf T}} 
\def\charge{{\sf C}} 
\def\pcp{p_{\text{PCP}}}
\def\TAB{T_{\text{AB}}}
\def\xiGL{\xi_{\text{GL}}}
\def\tauGL{\tau_{\text{GL}}}
\def\bxiGL{\bar{\xi}_{\text{GL}}}
\def\btauGL{\bar{\tau}_{\text{GL}}}
\def\text#1{\mbox{\tiny #1}}
\def\ket#1{\mbox{$\displaystyle\vert\,#1\,\rangle$}}

\def\nicefrac#1#2{\genfrac{}{}{}{1}{#1}{#2}} 
\def\Tr#1{\mbox{Tr}\big\{#1\big\}}
\def\der#1#2{\mbox{$\displaystyle\frac{d #1}{d #2}$}}
\begin{document}
\title{\large Kibble-Zurek Dynamics \& Statistics of Topological 
Defects in Chiral Superfluid \He\ Films}
\author{Noble Gluscevich}
\email{nglusc1@lsu.edu}
\affiliation{Hearne Institute of Theoretical Physics, 
             Louisiana State University,
	     Baton Rouge LA 70803, USA}
\author{J.A. Sauls}
\email{sauls@lsu.edu}
\affiliation{Hearne Institute of Theoretical Physics,
	     Louisiana State University,
	     Baton Rouge LA 70803, USA}
\date{\today}
\begin{abstract}
In equilibrium, confined films of superfluid $^3$He-A have the chiral axis, $\hat{\ns\vell}$, locked normal to the surface of the film. There are two degenerate ground states $\hat{\ns\vell}\;||\pm\hat\vz$.
However, for a temperature quench, i.e. cool down through the phase transition at a finite rate, causally disconnected regions of order parameter fluctuations develop and evolve into an inhomogeneous ordered phase that hosts both domain walls between time-reversed chiral phases as well as vortices with winding numbers $p\in\mathbb{Z}$. 
We present simulations based on a time-dependent generalization of Ginzburg-Landau theory for strong-coupling $^3$He that reveal both types of topological defects to be present following the temperature quench.
Results for the dynamics of vortices interacting with anti-vortices as well as domain walls are presented.
The vortex number density as a function of quench rate agrees well with the scaling predicted by Kibble and Zurek.
We also present results for the number distribution and compare with other theoretical models for full counting statistics of the topological defect density.
Finally, we present results for an asymmetry in the post-freeze-out populations of inequivalent vortex core structures that are characteristic of a chiral superfluid.
\end{abstract} 
\maketitle

\section{Introduction}
\vspace*{-5mm}

Superfluid \He\ is a paradigm for field theories describing spontaneous symmetry breaking and phase transitions in systems ranging from topological superconductors,\cite{miz18d,vol19} mass generation in particle physics,\cite{nam85,vol15,sau17} to cosmological phase transitions based on physics beyond the standard model.\cite{hin95,hin24}
The breadth of impact of the superfluid phases of \He\ originates from symmetry breaking at a second-order phase transition to superfluid phases that are BCS condensates belonging to a high dimensional representation of the maximal symmetry group,
$\G=\spin\times\orbital\times$ 
$\gauge\times\parity\times$
$\time\times\charge$.
The symmetry of the normal Fermi liquid phase of \He\ includes continuous symmetries under rotations ($\spin$) of the spin degrees of freedom of the Cooper pair condensate, rotations ($\orbital$) of the orbital degrees of freedom, and global gauge transformations ($\gauge$) corresponding to a uniform change in phase of the pair condensate. In addition, normal \He\ has discrete symmetries under parity ($\parity$), time reversal ($\time$) and fermion charge conjugation parity ($\charge$). The latter is emergent as a symmetry of the low-energy effective field theory of the normal Fermi liquid under particle $\leftrightarrow$ hole conversion.
The field theory describing the broken symmetry phases of \He, as well as the space-time dynamics of the Bosonic fields describing them, is constrained by the symmetry group ${\sf G}$, and has been extended to incorporate strong-coupling corrections to weak-coupling BCS theory.~\cite{wim16,sau17,miz18c,reg20} A review of the time-dependent Ginzburg-Landau (TDGL) field theory for dynamics and phase transitions in superfluid \He\ is provided in Ref.~\onlinecite{hin24}.

The breaking of one or more continuous symmetries has consequences for the vacuum state, the generation of low-energy Bosonic excitations as Nambu-Goldstone modes, and the emergence of topologically stable defects that can become embedded in the otherwise homogeneous vacuum.
In the case of superfluid \Hefour\ global $\gauge$ symmetry is broken by the macroscopic wave function of the scalar Bose-Einstein condensate. 
The resulting vacuum state is degenerate, defined by the possible phases of the order parameter, i.e. the manifold $\cM=S^1$. The latter determines the fundamental homotopy group defining the classes of topologically stable line defects, $\pi_1(S^1)=\mathbb{Z}$, identified with the discrete winding numbers of quantized phase vortices, $N_{\cC}=\frac{1}{2\pi}\oint_{\cC}\,\grad\varphi(\vr)\cdot d\vl\in\mathbb{Z}$.

\vspace{-10mm}
\subsubsection{Kibble-Zurek Mechanism}
\vspace{-3mm}

In symmetry breaking phase transitions, topological defects can form in the symmetry-broken phase as residuals of the high-symmetry phase. The number of such defects is controlled, in part, by the rate at which the phase transition takes place, i.e. the ``quench rate'', $1/\tau_Q$. 
In the cosmological scenario considered by Kibble continuous symmetries of a quantum field theory are assumed to be broken in the early universe during rapid cooling through a second order phase transition. Near the transition long-lived fluctuations of the order paramter just above the transition cannot equilibrate to the changing temperature, leading to causally disconnected regions of the broken symmetry phase that grow in size. At later times these distinct regions, or ``superfluid patches'',  each belonging to an element of the degeneracy manifold, $\cM$, connect to form a macroscopic inhomogeneous ordered phase with topologically stable defects - monopoles (point defects), cosmic strings and vortices (line defects), and domain walls - embedded in one or more broken symmetry states.~\cite{kib76,kib80}

Motivated by Kibble's analysis for cosmological phase transitions, Zurek considered defect formation in condensed matter systems undergoing a second order phase transition, particularly the Helium liquids.~\cite{zur85,zur93} He also formulated a scaling relation relating the number of topological line defects to the quench rate and critical exponents that define the correlation length and order parameter relaxation time for temperatures close to the transtion.
Since then topological defect creation has been observed in a variety of condensed matter systems via a temperature or pressure quench through a second order phase transition, ranging from superfluid \He,~\cite{bau96,ruu96,ruu98,elt00} superconductors,~\cite{man03} nematic liquid crystals,~\cite{chu91,bow94,hen94} \He\ confined in nematic aerogel,~\cite{aut16,rys21} to cold atomic gases.~\cite{wei08,lam13,lee24}

\vspace{-8mm}
\subsubsection{Scaling Relation for the Defect Density}
\vspace{-3mm}

For a system in thermodynamic equilibrium near a second-order phase transition to a broken symmetry state fluctuations of the ordered phase with characteristic size $\xi(T)$, the \emph{correlation length}, persist with a lifetime, $\tau(T)$, both diverging as $T\rightarrow T_c$,
\begin{equation}
\xi(T)=\frac{\xi_0}{\left\vert T/T_c - 1\right\vert^\nu}
\,,
\end{equation}
where $\xi_0$ is the zero-temperature correlation length and $\nu$ is the corresponding critical exponent.
Similarly, 
\begin{equation}\label{eq-tau_vs_T}
\tau(T) = \frac{\tau_0}{\left\vert T/T_c - 1\right\vert^{z\nu}}
\,,
\end{equation}
where $z$ is the dynamical exponent.
Note that $\xi_0$ is the healing length of the fully developed order parameter due to a perturbation that locally destroys long-range order, e.g. the singularity at the core of a quantized vortex line in superfluid \He, and $\tau_0$ is the characteristic time for the order parameter to recover from an impulse that destroys the order parameter, e.g. the pair formation timescale in superfluid \He.
When the temperature is driven through the transition at a finite rate, 
\begin{equation}
\frac{1}{\tau_Q} = -\frac{1}{T_c}\,\der{T}{t}\Big\vert_{T_c}
\,,
\end{equation}
there is a temperature window near $T_c$ where the lifetime of the fluctuations of the ordered phase is comparable to or longer than the time to cross the temperature window near $T_c$. 
This is \emph{critical slowing down} near a phase transition. As a result causally disconnected regions of the ordered phase form and grow in size and amplitude just below $T_c$.
These ``patches'' of the ordered phase, each belonging to a distinct element of the degeneracy manifold, merge at the \emph{freeze-out} time, $\hat{t}=\tau(T(\hat{t}))$,\cite{zur85,zur96}
\begin{equation}
\hat{t} = \tau_Q\,\left(\frac{\tau_0}{\tau_Q}\right)^{\frac{1}{z\nu+1}}
\,.
\end{equation}
Topologically stable defects form when local currents cannot anneal the ordered phase to a uniform condensate in any finite region of space. 
The typical size of the patches of correlated order parameter at freeze-out is then given by 
\begin{equation}
\hat{\xi}\equiv\xi\left(T(\hat{t})\right)=\frac{\xi_0}{(\tau_0/\tau_Q)^{\frac{\nu}{z\nu+1}}} 
\,.
\end{equation}
Thus, the density of topological defects scales inversely with a power of $\hat{\xi}$. For line defects in $d=3$ space dimensions, or point defects in $d=2$, the mean areal density of topological line or point defects is 
\begin{equation}
\frac{\langle n \rangle}{L^2}
= f\,\hat{\xi}^{-2}
= \frac{f}{\xi_0^2}\,
  \left(\frac{\tau_0}{\tau_Q}\right)^{\beta} 
\,,
\label{eq-KZ_scaling}
\end{equation}
where $\beta=2\nu/(1 + z \nu)$ and $f$ is the fraction of freeze-out regions that host an embedded topological defect, and $L$ is the linear dimension of the system. This scaling relation for the mean number, $\langle n\rangle$, of defects is a key prediction of the Kibble-Zurek mechanism (KZM). Note that parameter $f$ depends on the specifics of the dynamics of the nonequilibrium order paramter fluctuations under the temperature quench.

Large scale numerical simulations of vortex string generation in $d=3$ space dimensions were reported for the second-order phase transition of a $\gauge$ scalar field with Langevin noise and damping by Antunes et al.~\cite{ant99} Their results confirmed KZM scaling, albeit with an exponent $\beta$ that varied between $0.43$ to $0.57$ depending on the numerical criterion for identifying the freeze-out time. Also, notable is the deviation of the exponent from the mean-field prediction of $\beta_{\text{MF}}=0.5$. 

\vspace{-5mm}
\subsubsection{Counting Statistics}
\vspace{-3mm}

Equation~\eqref{eq-KZ_scaling} corresponds to the first moment of the probability distribution, $\cP(n)$, for the generation of $n$ topological defects after a quench.
A recent interpretation of the KZM provides additional insight into the statistics of topological defect generation by a quench.~\cite{del18,gom20}
These authors argue that the merging of causally disconnected regions at freeze-out generates a topological defect with probability $p$ that is independent of other defects as a result of spatial separation at the freeze-out time.
Thus, for a lattice model of a field theory with $N$ possible sites for a defect the number of defects formed during a quench is expected to follow a binomial distribution,
\begin{equation}
\cP_{\text{N}}(n) = \frac{N!}{n!\,(N-n)!}\,p^n\,(1-p)^{N-n}
\,.
\end{equation}
A key feature of the distribution is that all cumulants, $\kappa_i$ for $i\in\{1,2,3,\ldots\}$,~\footnote{The first three cumulants are the mean, $\kappa_1 = \langle n \rangle$,  variance, $\kappa_2 = \langle (n -\langle n \rangle)^2\rangle$,  and skewness, $\kappa_3 = \langle (n -\langle n \rangle)^3\rangle$. For a discussion of the higher order cumulants see Ref.~\onlinecite{gom20}.} scale with the quench rate with the \emph{same} exponent, i.e. $\kappa_i \propto \left(1/\tau_Q\right)^{\beta}$.
Such an uncorrelated probability distribution for the formation of $n$ defects was confirmed by numerical simulations of defects (``kinks'') in a $d=1$ chain of atoms with nearest neighbor interactions that exhibit a structural phase transition from a linear chain at high temperature to a doubly-degenerate zig-zag phase below $T_c$.
In higher spatial dimensions counting defects is more complicated, and the probability distribution becomes a \textit{Poisson} binomial. Nevertheless, if this distribution describes the freeze-out of topological defects by the KZM it too predicts that all cumulants obey Kibble-Zurek scaling with the same exponent.~\cite{gom20}

\vspace{-5mm}
\subsubsection{KZM in Chiral Superfluid \He\ Films}
\vspace{-3mm}

Here we report theoretical predictions for the dynamics and statistics of topological defects in films of the \emph{chiral phase of superfluid \He} following a temperature quench through $T_c$.
The chiral phase is the ground state of liquid \He\ at all pressures for films with thicknesses $D\le D_{c_2}=9\xi_0\approx$ $0.69\,\mu\mbox{m}$ at $0\,\mbox{bar}$ ($0.14\,\mu\mbox{m}$ at $p=34\,\mbox{bar}$). 
The stability of the chiral A-phase for all pressures below solidification and temperatures below the bulk superfluid transition in films of thickness $D=80\,\mbox{nm}$ is reported in Ref.~\onlinecite{hei25}. The authors also control surface boundary scattering and surface pair-breaking. The authors show that surface pair breaking is almost completely suppressed by plating cavity surfaces with a few atomic layers of \Hefour. These are the confinement and boundary conditions we assume in our simulations.
We also assume the spin quantization axis is fixed normal to the film, $\hat{\vd}=+\hat{\vz}$ in zero magnetic field (see Sec. II.2).
The chiral phase is then described by a two-dimensional complex vector order parameter, $\vA(\vr,t)$, confined to the plane of the film. The ground state breaks spin- and orbital rotation symmetries in conjunction with global $\gauge$ symmetry, as well as discrete symmetries of parity ($\parity$) and time-reversal ($\time$).
The residual symmetry group of the chiral phase includes spin-rotations about the normal to the plane of the film, $\point{U(1)}{L$_z$-N}$ gauge-orbit symmetry,~\footnote{Rotations about the chiral axis reduce to an overall phase factor which can be undone with a gauge transformation.} and the discrete $\mathbb{Z}_2$ symmetry corresponding to the product of time-reversal and mirror reflection in a plane containing the $z$-axis. 
In addition to the degeneracy manifold defined by the global phase, $\cM=S^1$, the breaking of time-reversal and mirror symmetries implies the ground state is doubly degenerate with $\vA_{\pm} = A_0\,e^{i\vartheta}\,\left(\hat\vx \pm i \hat\vy\right)/\sqrt{2}$. These two distinct chiral phases correspond to condensates of Cooper pairs with ground-state angular momentum $\langle L_z \rangle = \pm (N/2)\hbar$ for a film with $N$ \He\ atoms.~\cite{sau11} Thus, in addition to phase vortices classified by $\pi_1(S^1)=\mathbb{Z}$, the \He\ films support topologically stable \emph{domain walls} separating the two distinct chiral ground states. 
Furthermore, broken time-reversal symmetry implies that vortices and anti-vortices in a fixed chiral ground state are non-degenerate defects with inequivalent internal structure.~\cite{tok90,sau09}
It is a priori unknown how domain walls and their dynamics impacts the generation and dynamics of vortices and anti-vortices.
Thus, it is an open question whether or not KZM scaling differs from that of $\gauge$ scalar field theory, and furthermore whether the statistics of defect generation differs for vortices and anti-vortices, or if the statistics is Poisson binomial.  

\vspace*{-5mm}
\section{TDGL Theory for $^3$He}\label{sec-TDGL}
\vspace*{-3mm}

Time-dependent generalizations of Ginzburg-Landau theory were developed by a number of authors to describe the dynamics of superconductors,~\cite{abr66,kop02} quench dynamics and normal-superfluid boundary propagation in $\point{U(1)}{}$ superfluids,~\cite{ara99,kop99} as well as order parameter dynamics of superfluid \He.~\cite{kle78,sau17}
In what follows we use the strong-coupling GL theory~\cite{wim16,sau17} developed for superfluid \He\ to study the nonequilibrium dynamics of the order parameter, including the generation and dynamics of topological defects in thin films of chiral superfluid \He\ following a temperature quench through the phase transition temperature $T_c$.
We start from the formulation of TDGL field theory in Ref.~\onlinecite{hin24} for the dynamics of the order parameter in \He, including its coupling to thermal noise from the underlying fermionic environment described by a stochastic Langevin source term and its associated damping.

\vspace*{-5mm}
\subsubsection{Ginzburg-Landau Functional}\label{sec-GL_functional}
\vspace*{-3mm}

The superfluid phases of \He\ belong to the spin-triplet ($S=1$), p-wave ($L=1$) manifold of pairing states defined by the pair condensate amplitude, $\langle\psi_{\vp,a}\psi_{-\vp,b}\rangle$, where $\vp$ is the relative momentum of a pair of orbiting fermions, and $a,b\in\{\uparrow\,,\downarrow\}$ are the spin projections of the pair of fermions.
The corresponding mean-field pairing energies are $\Delta_{ab}(\vp)=g\,\langle\psi_{\vp,a}\psi_{-\vp,b}\rangle$, where $g$ the attractive pairing interaction in the triplet, p-wave channel. These amplitudes are the elements of a $2\times 2$ matrix, $\widehat\Delta(\vp)=i\vec{\sigma}\sigma_y\cdot\vec{d}(\vp)$, where $i\vec\sigma\sigma_y$ are the three symmetric $2\times 2$ Pauli matrices, and $\vec{d}(\vp)=\sum_{\alpha}\,d_{\alpha}(\vp)\,\vec{e}_{\alpha}$ is a vector under spin rotations.
The components $d_{\alpha}(\vp)$ are functions of the p-wave basis functions for momenta restricted to the Fermi surface. Thus, the general form for the manifold of p-wave, spin-triplet order paramters, $d_{\alpha}(\vp) = \sum_{i}\,A_{\alpha i}\,\hat\vp_i$, can be expressed in terms of nine complex amplitudes, $A_{\alpha i}$, that form the elements of a $3\times 3$ matrix that transform as the vector representations of $\point{SO(3)}{}$ \emph{separately} under spin and  orbital rotations.\footnote{The spin basis, $\{\vec{e}_{\alpha} | \alpha=x',y',z'\}$, is an orthogonal triad of unit vectors in spin space, while the orbital basis functions, $\hat\vp_i=\ve_i\cdot\hat\vp$, are the direction cosines of the orbital momentum, $\vp$, defined on the Fermi surface, and referenced to the unit vectors $\{\ve_i | i=x,y,z\}$ in orbital space.} These transformations, and the symmetry constraints imposed by the maximal symmetry group, 
\begin{equation}
\G=\spin\times\orbital\times\gauge\times\parity\times\time\times\charge
\,,
\label{eq:bulk_symmetry}
\end{equation}
define the effective field theory governing the superfluid phases of \He\ which is the basis for our study of the dynamics of quasi-two-dimensional films of \He.

The space-time evolution of the order parameter describing Cooper pairs in bulk $^3$He, $A_{\alpha i}(\vr,t)$, is governed by field equations obtained from the Lagrangian, $L=\int\,dV\,\cL[A,\dot A,\partial A]$, where the Lagrangian density is
\begin{eqnarray}\label{eq-Lagrangian}
\mathcal{L} 
&=& 
\mu\,\Tr{\dot A\dot A^{\dag}}
-
\alpha\,\Tr{A\,A^{\dag}}
\nonumber\\
&-&
\sum_{p=1}^{5}\beta_{p\,}u_{p}(A)
- 
\sum_{m=1}^{3}\,K_{m}\,v_{m}(\partial A)
\,,
\end{eqnarray}
where $\dot A = \partial_t A$ and $\Tr{A\,A^{\dag}}$ is the second-order bulk invariant that controls the phase transition to the broken symmetry phases; for $\alpha >0$ the equilibrium state is the un-broken normal state, while for $\alpha<0$ the equilibrium state spontaneously breaks one or more symmetries of $\G$ and develops a macroscopic Cooper pair amplitude, $A$. 
The broken symmetry equilibrium state is determined by the five linearly independent fourth-order invariants, 
\begin{eqnarray}
u_1 
&\ns=\ns& 
|\Tr{A A^T}|^2 
\,,\,
u_2 = \Tr{A A^\dagger}^2 
\,,\,
u_3 = \Tr{A A^T A^* A^\dagger} 
\,,
\nonumber\\
u_4 
&\ns=\ns&
\Tr{A A^\dagger A A^\dagger} 
\,,\,
u_5 = \Tr{A A^\dagger A^* A^T }
\,,
\end{eqnarray}
where $A^T$ ($A^\dagger$) is the transpose (adjoint) of $A$.
The space-time dynamics of the order parameter is generated in part by the spatial and temporal invariants,\footnote{Violation of particle-hole symmetry by the parent Fermi-liquid allows for an additional invariant in the Lagrangian that is first-order in $\partial_t A$, i.e. $K_{\Gamma} = i\Gamma \left[\Tr{\dot{A} A^\dagger} - \Tr{A \dot{A}^\dagger}\right]$. This C-violating term is small so we omit it in the dynamical simulations that follow. This term in the Lagrangian is unrelated to the dissipative term that is first order in $\partial_t A$ that arises from coupling of the thermal bath of fermionic excitations to nonequilibrium states of the Bosonic field discussed in Sec.~\ref{sec-TDGL+dissipation}.}
\begin{eqnarray}
v_0 &=& \partial_t A_{\alpha i}\partial_t A_{\alpha i}^*
\,,\quad
v_1 = \partial_k A_{\alpha j}\partial_k A^*_{\alpha j}
\,,
\nonumber\\
v_2 &=& \partial_j A_{\alpha j}\partial_k A^*_{\alpha k} 
\,,\quad
v_3 = \partial_k A_{\alpha j}\partial_j A^*_{\alpha k}
\,.
\end{eqnarray}
The coefficients, $\alpha$, $\beta_p$, $\mu$ and $K_m$ that weight the contributions to the effective Lagrangian have been calculated from the microscopic theory of superfluid \He. In particular the bulk parameters are, 
\begin{equation}
\alpha = \frac{1}{3}N(0)\left(T/T_c -1\right) 
\,,\quad
\beta_p = \beta_0\left(b_p^{\text{wc}} + \frac{T}{T_c} b_p^{\text{sc}}\right)\,,
\label{eq-alpha+beta_parameters}
\end{equation}
for $p\in\{1,\ldots, 5\}$. Note that $N(0)=m^{*}k_f/(2\pi^{2}\hbar^{2})$ is the normal-state density of states at the Fermi surface, $k_f$ is the Fermi wavevector, $m^{*}$ is the quasiparticle effective mass, and $T_c$ is the bulk superfluid transition temperature. All are known functions of pressure.
In the weak-coupling limit the $\beta_p$ parameters are determined by pressure-independent 
ratios and a pressure-dependent scale factor, 
\begin{equation}
\{ b_p^{\text{wc}} \} = (-1, 2, 2, 2, -2)
\,,\quad 
\beta_0=\frac{7\zeta(3)}{80\pi^2} \frac{N(0)}{3 (\kb T_c)^2}
\,.
\end{equation}

The strong-coupling corrections to the $\beta_p$ parameters, $b_p^{sc}$, were calculated based on the leading order corrections to weak-coupling BCS theory starting from the formalism of Rainer and Serene.~\cite{rai76,sau81b,sau81c,wim19}
The more recent development of the strong-coupling free-energy functional is now able to predict the stability of the A-phase above the polycritical pressure, as well as the temperature dependence of the excitation gap, thermodynamic potential and heat capacity of the B-phase down to low temperatures. The theoretically predicted AB transition line based on the Luttinger-Ward functional is in excellent agreement with the experimental AB transition as a function of pressure.~\cite{wim19}
Another key development of the strong-coupling GL functional was inclusion of the temperature-dependent scaling of the strong-coupling $\beta_p$ parameters in Eq.~\eqref{eq-alpha+beta_parameters}. 
The reduction in the strong-coupling contribution follows from microscopic strong-coupling theory and was introduced in Ref.~\onlinecite{wim16} where it was shown that the temperature and pressure dependence of the strong-coupling $\beta_p$ parameters predicts the AB transition transition line to high accuracy.
This is equally true for the most recent set of strong-coupling $\beta_p$ parameters based on leading order corrections to weak-coupling theory.~\cite{wim19}
This temperature scaling extends the predictive capabilities of the GL theory to temperatures below $T_{\text{AB}}$ for pressures above the polycritical pressure,~\cite{reg20} and is essential for developing TDGL theory to study order parameter dynamics in strong-coupling superfluid \He.~\cite{hin24} 
In this report the set of strong-coupling $b_p^{\text{sc}}$ parameters are computed from $4^{\textrm{th}}$-order polynomial fits to the computed values reported in Ref.~\onlinecite{wim19}.

The coefficients of the kinetic and gradient terms are given to high accuracy by their weak-coupling values,
\begin{equation}
\hspace*{-1mm}
\mu=\frac{7\zeta(3)}{12}\,N(0)\,\tau_0^2
\,,
K_1 = K_2 = K_3 =\frac{7\zeta(3)}{60}\,N(0)\,\xi_0^2
\,,
\label{eq-mu+K_parameters}
\end{equation}
where $\tau_0=\nicefrac{\hbar}{2\pi\kb T_c}$ is the zero-temperature Cooper pair formation timescale and $\xi_0 = v_f\tau_0$ is the corresponding Cooper pair coherence length.
These coefficients define the temperature-dependent GL pair formation time and coherence length,\footnote{Note that temperature-dependent scaling of $\tauGL(T)$ differs from that in Eq.~\eqref{eq-tau_vs_T} because the latter defines the time scale for pair fluctuations in the critical region in which the order parameter has not yet developed.}
\begin{eqnarray}
\tauGL(T) 
&=& 
\frac{\bar{\tau}_{\text{GL}}}{\left(1-T/T_c\right)^{\tinyonehalf}}
\,,\quad
\bar{\tau}_{\text{GL}}=\sqrt{\frac{7\zeta(3)}{12}}\tau_0
\,,
\\
\xiGL(T) 
&=& 
\frac{\bar{\xi}_{\text{GL}}}{\left(1-T/T_c\right)^{\tinyonehalf}}
\,,\quad
\bar{\xi}_{\text{GL}}=\sqrt{\frac{7\zeta(3)}{20}}\xi_0
\,.
\end{eqnarray}

\subsubsection{TDGL Field Theory for \He\ Films}\label{sec-TDGL+dissipation}

Bulk \Hea\ is the equilibrium phase at pressures above $\pcp=21\,\mbox{bar}$ and temperatures $\TAB\le T\le T_c$ (c.f. Fig. 1 in Ref.\onlinecite{reg20}). Its stability results from corrections to weak-coupling BCS theory that favor equal-spin pairing (ESP) states and become sufficiently large at high pressures to stabilize an ESP state that spontaneously breaks time-reversal symmetry with an order parameter of the form,
\begin{equation}
A_{\alpha i}^{\text{A}} 
= 
\Delta_{\text{A}}\,\hat{\vd}_{\alpha}\,
\left(\hat{\vm}_i \pm i\,\hat{\vn}_i\right)/\sqrt{2}
\,,
\end{equation}
where $\Delta_{\text{A}}$ is the equilibrium amplitude of the bulk A-phase order parameter, $\hat\vd$ is a real unit vector in spin space along which the A-phase Cooper pairs have zero spin projection. Alternatively, the pairs are in an equal-spin pairing (ESP) state with equal amplitudes for pairs in the states $\ket{\uparrow\uparrow}$ and $\ket{\downarrow\downarrow}$ confined in the plane \emph{perpendicular} to $\hat\vd$.
Orthonormal unit vectors, $\hat\vm$ and $\hat\vn$, combined with a relative phase of $\pm\pi/2$, define the orbital motion of the A-phase Cooper pairs with angular momentum $\pm\hbar$ per Cooper pair along the axis $\hat{\ns\vell}=\hat\vm\times\hat\vn$. This pseudo vector highlights broken mirror and time-reversal symmetry by the A-phase. 

In films or slabs of confined \He\ with specularly reflecting surfaces and thickness $D\lesssim 9\;\xi_0$ boundary scattering completely suppresses out-of-plane orbital states $\propto\hat{p}_z$, while strong-coupling effects favor the ESP A-phase with $\hat{\ns\vell}=\pm\hat\vz$ over the planar phase. As a result the A-phase is stabilized for all pressures below melting pressure, and temperatures below the bulk transition temperature.
The spin and orbital broken symmetry directions, $\hat\vd$ and $\hat{\ns\vell}$, are coupled by the nuclear dipole-dipole interaction, $u_{\text{D}}=-g_{\text{D}}\,\Delta_{\text{A}}^2\,(\hat\vd\cdot\hat{\ns\vell})^2$, with $g_{\text{D}} >0$, which \emph{locks} $\hat\vd||\pm\hat{\ns\vell}$.
Thus, for \He\ films in the zero-field we restrict our analysis to the dipole-locked case with $\hat\vd\parallel\hat\vz$.~\footnote{In a magnetic field with $\vB||\pm\vell$ and $B\ge B_c\simeq 2.5\,\mbox{mT}$ the Zeeman energy dominates, is minimized for $\hat\vd\perp\vB$, which ``unlocks'' $\hat\vd$ from $\hat{\ns\vell}$.}
The general form of the order parameter then reduces to a two-dimensional complex vector for the in-plane orbital order parameter, $A_{\alpha i}\rightarrow \hat\vz_\alpha\,A_i$, with
\begin{equation}
\vA(\vr,t) = A_x(\vr,t)\ve_x + A_y(\vr,t)\ve_y
\,,
\end{equation}
where the Cartesian basis vectors, $\ve_x$ and $\ve_y$, are real and orthonormal, and $A_x$ and $A_y$ are complex functions of space-time.
The corresponding Lagrangian density for dipole-locked \He\ films reduces to
\begin{eqnarray}\label{eq-Lagrangian-film}
\mathcal{L} 
=
\mu\,\dot A_i\,\dot A_i^*
-
\alpha\,A_i\,A_i^*
-
\beta_{13}(A_i A_i)(A_j^* A_j^*)
-
\beta_{245}(A_i A_i^*)^2
\nonumber\\
\ns-\ns 
K_1\,(\partial_i A_j)(\partial_i A_j^*)
\ns-\ns 
K_2\,(\partial_i A_i)(\partial_j A_j^*)
\ns-\ns 
K_3\,(\partial_i A_j)(\partial_j A_i^*)
,\hspace*{2mm}
\end{eqnarray}
where $\beta_{13}\equiv\beta_1+\beta_3$ and $\beta_{245}=\beta_2+\beta_4+\beta_5$ determine the possible ground states of the \He\ film for $T<T_c$ ($\alpha<0$).\footnote{This Lagrangian and its extension to p-wave superconductors with tetragonal symmetry was used to investigate the collective modes of the chiral p-wave model for \sro\ in Ref.\onlinecite{sau15}.} 
If $\beta_{13} < 0$ and $\beta_{13}+\beta_{245}>0$ the ground state is an in-plane polar phase with $\vA^{\text{P}} = \Delta_{\text{P}}\,(\cos\chi\ve_x+\sin\chi\ve_y)$ where $\chi\in\{0,2\pi\}$ defines the orientation and continuous degeneracy of the in-plane polar state.
However, for $\beta_{13} > 0$ and $\beta_{245}>0$ the ground state is 
the doubly-degenerate chiral state with 
$\vA^{\text{A}}_{\pm}=\Delta_{\text{A}}\,(\ve_x \pm i\ve_y)/\sqrt{2}$, and
$\Delta_{\text{A}}=\sqrt{|\alpha(T)|/2\beta_{245}}$.

\begin{figure*}
\centering
\includegraphics[width=0.9\textwidth]{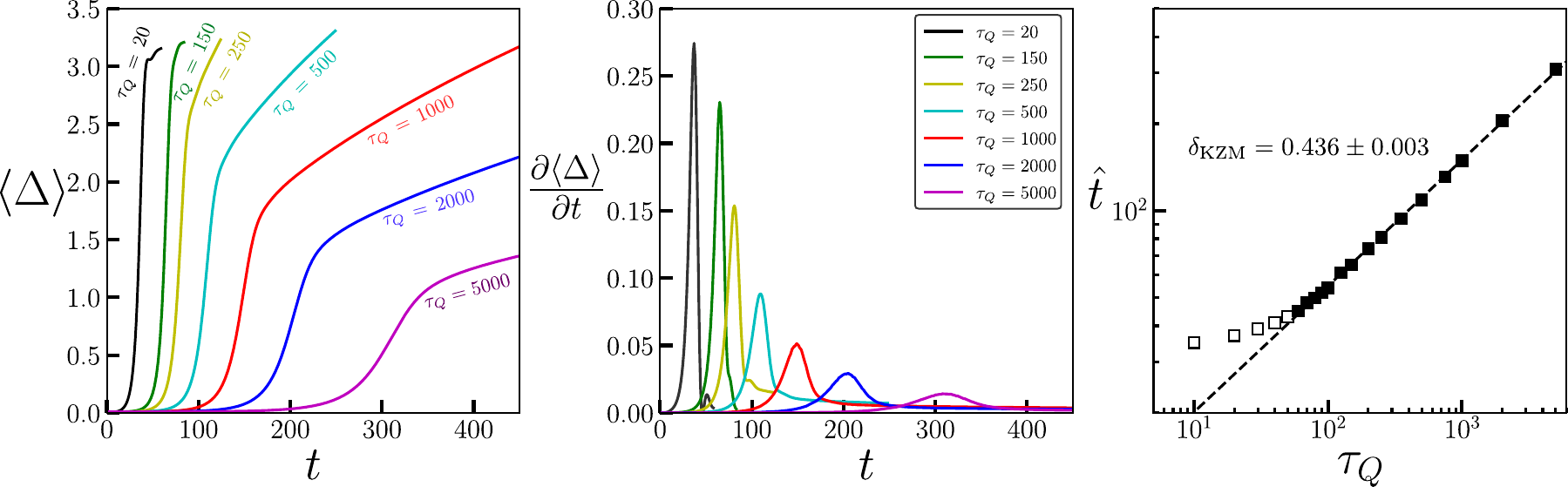}
\caption{
Left panel: Spatially averaged order parameter (Eq.~\eqref{eq-Average_Delta}, in units $k_B T_c$) as a function of time (in units $\bar{\tau}_{\text{GL}}$) after crossing $T_c$ for quench times: $\tau_Q/\btauGL=20,150,250,500,1000,2000,5000$.
Center panel: The freeze-out time is identified with the maximum of $\partial_t\langle\Delta\rangle$. 
Right panel: The resulting freeze-out time, $\hat{t}$, as a function of quench time. Kibble-Zurek scaling: $\hat{t}\sim\tau_Q^{\delta_{\text{KZM}}}$ is obeyed over two orders of magnitude in $\tau_Q$ with exponent $\delta_{\text{KZM}}=0.436$. Unfilled data points exhibit saturation and are not included in the fit.
}
\label{fig:nonequilibrium}
\end{figure*}

\vspace*{-5mm}
\subsubsection{Langevin Noise and Dissipation}
\vspace*{-3mm}

The Euler-Lagrange equations for the complex vector field lead to the equations for the space-time dynamics of the order parameter. 
In addition to the conservative dynamics of the bosonic field theory we follow Refs.~\onlinecite{ant99,hin24} and include the coupling of the bosonic excitations to the underlying thermal bath via a stochastic Langevin noise source, $\zeta_i(\vr,t)$, with time average $\langle \zeta_i(\vr,t)\rangle=0$, and Gaussian correlator,
\begin{equation}
\langle \zeta_i(\vr,t)\zeta^*_i(\vr',t') \rangle 
=
2\gamma\kb T \delta_{ij}\delta(\vr-\vr')\delta(t-t')
\,.
\end{equation}
The parameter $\gamma$ plays an important role as it leads to damping of space-time fluctuations of the bosonic field via an additional dissipative time-derivative term, $\gamma\dot A_i$, that is characteristic of Langevin dynamics; 
$\gamma$ is constant in the ``gapless'' regime near $T_c$, $|\Delta| \ll \pi k_B T_c$, given by
\begin{equation}
\gamma = \frac{\hbar \pi N(0)}{48 k_B T_c}.
\end{equation}
Thus, the set of dynamical equations including the damping and Langevin noise source terms are,
\begin{eqnarray}\label{eq-TDGL_films}
&\mu\ddot{A}_i
+
\gamma\dot{A}_i
+
\alpha(T) A_i
-
K_1\partial^2 A_i
-
K_{23}\partial_i(\partial_j A_j)
\qquad\qquad
&
\\
&+
2\beta_{13}(A_j A_j)\,A_i^*
+
2\beta_{245}(A_j A_j^*)\,A_i
\ns=\ns 
\zeta_i(\vr, t)
\,,
\,i\in\{x,y\}
\,,
&
\nonumber
\end{eqnarray}
where $K_{23}=K_2+K_3$.
Strong-coupling calculations of the free energies of the bulk superfluid phases, including the $\beta_p$ parameters,~\cite{wim19} predict the bulk AB-transtion line,~\cite{wim16,wim19} as well as the metastability of A-core vortices at high pressures for rotating \Heb.~\cite{reg20,ran24} The theory also predicts that $\beta_{13} > 0$ for all pressures, implying that the dipole-locked A-phase is the ground state for specular \He\ films of thickness $D\lesssim 9\xi_0$ over the entire pressure range below the bulk $T_c$.
NMR and torsional oscillator measurements confirm that the A-phase is the ground state of \He\ confined in slabs of thickness $D<D_{c_2}(T)$ at low pressures and temperatures.~\cite{lev13,zhe17b}

Thus, in the dynamical simulations of non-equilibrium phase transitions via a temperature quench we expect topological defects to form and evolve within the chiral A phase. Thus, for analysis and visualization it is convenient to express the space-time evolution in terms of the chiral basis,
\begin{equation}
\vA(\vr,t) = A_+(\vr,t)\ve_+ + A_-(\vr,t)\ve_-
\,,
\end{equation}
where $\ve_{\pm}=(\ve_x \pm i \ve_y)/\sqrt{2}$ which satisfy the orthonormality conditions, $\ve_m\cdot\ve_n^* = \delta_{m,n}$ for $m,n\in\{+,-\}$. The transformation for the order parameter components from the Cartesian to chiral basis is $A_\pm=(A_x\mp iA_y)/\sqrt{2}$.
 
\vspace*{-5mm}
\section{Numerical Methods and Results}
\vspace*{-3mm}

To study the dynamics of the order parameter for \He\ films subject to a temperature quench, the formation of topological defects and their later stage dynamics we solve Eqs.~\eqref{eq-TDGL_films} with the material parameters described in Sec.~\ref{sec-GL_functional}.
Simulations are performed on a computational domain of size, $L_x=150\,\bxiGL\times L_y=150\,\bxiGL$, with periodic boundary conditions enforced on the order parameter.
The equations of motion are solved with a finite difference explicit-time method, with grid spacings, $\delta x =\delta y = 0.5\bxiGL$ and $\delta t=0.025\btauGL$ \cite{strikwerda04}. For finer scale ``snapshots'' of the order parameter configuration showcasing vortex structure and mass current density, we use $\delta x = \delta y = 0.25\bxiGL$. At each time step, the Gaussian noise with random phase is implemented at every grid point, where the Gaussian noise is generated by the Box-Muller transform and multiplying by a random number on the unit circle in the complex plane.\cite{box58} 

To test Kibble-Zurek (KZ) scaling for the number of topological defects that form we identify the KZ ``freeze-out time'' from the onset of rapid growth of the spatially averaged order parameter,
\begin{equation}\label{eq-Average_Delta}
\langle\Delta\rangle_t\equiv\int\frac{d^2r}{L_xL_y}\,
\sqrt{\vA(\vr,t)\cdot\vA(\vr,t)^*}
\,,
\end{equation}
as $T(t)$ drops below $T_c$. 
Fig.~\ref{fig:nonequilibrium} shows the evolution of the spatially averaged order parameter as a function of time after the temperature drops below $T_c$ for quench times ranging from $\tau_Q=20\,\btauGL$ to $\tau_Q=5000\,\btauGL$, averaged over $N=20000$ random noise realizations for $\tau_Q\le 1000\,\btauGL$ and $N=375$ for $\tau_Q/\btauGL=2000,5000$.
In contrast to adiabatic growth under quasi-equilibrium conditions, which follows a square root trajectory until the final temperature, for a finite quench rate there are two regimes of order parameter growth: \emph{impulse} 
response, followed by quasi-adiabatic growth.
In the impulse region near $T_c$, the order parameter growth timescale is relatively long, and the order parameter is not able to respond to changes in temperature. At later times the order parameter responds quickly to changes in temperature, and achieves a near-equilibrium value, reduced slightly by the number of topological defects that remain.
The boundary between these two regimes is identified as the freeze-out time,
similar to that in Ref.~\onlinecite{ant99}.
Operationally the freeze-out time, $\hat{t}$, is identified as the point of maximum slope of $\langle\Delta\rangle_t$, shown as the peak time in the center panel of Fig.~\ref{fig:nonequilibrium}.  
With this definition we find that $\hat{t}$ scales with the quench time with KZ exponent $\delta_{\text{KZM}}=0.436\pm 0.003$, an exponent close to that reported in Ref.~\onlinecite{ant99} for vortex string formation following a quench for the 3D $\point{U(1)}{}$ theory. Note also that the KZ scaling exponent is below the GL mean field prediction of $\delta_{\text{GL}}=1/2$.
The timescale for order parameter growth is also relevant to the nucleation of the B phase from the metastable A phase following local energy deposition that locally destroys superfluidity. Subsequent evolution of the order parameter can evolve into either B or A phase.~\cite{hin24}

\begin{figure}
\centering
\includegraphics[width=0.99\linewidth]{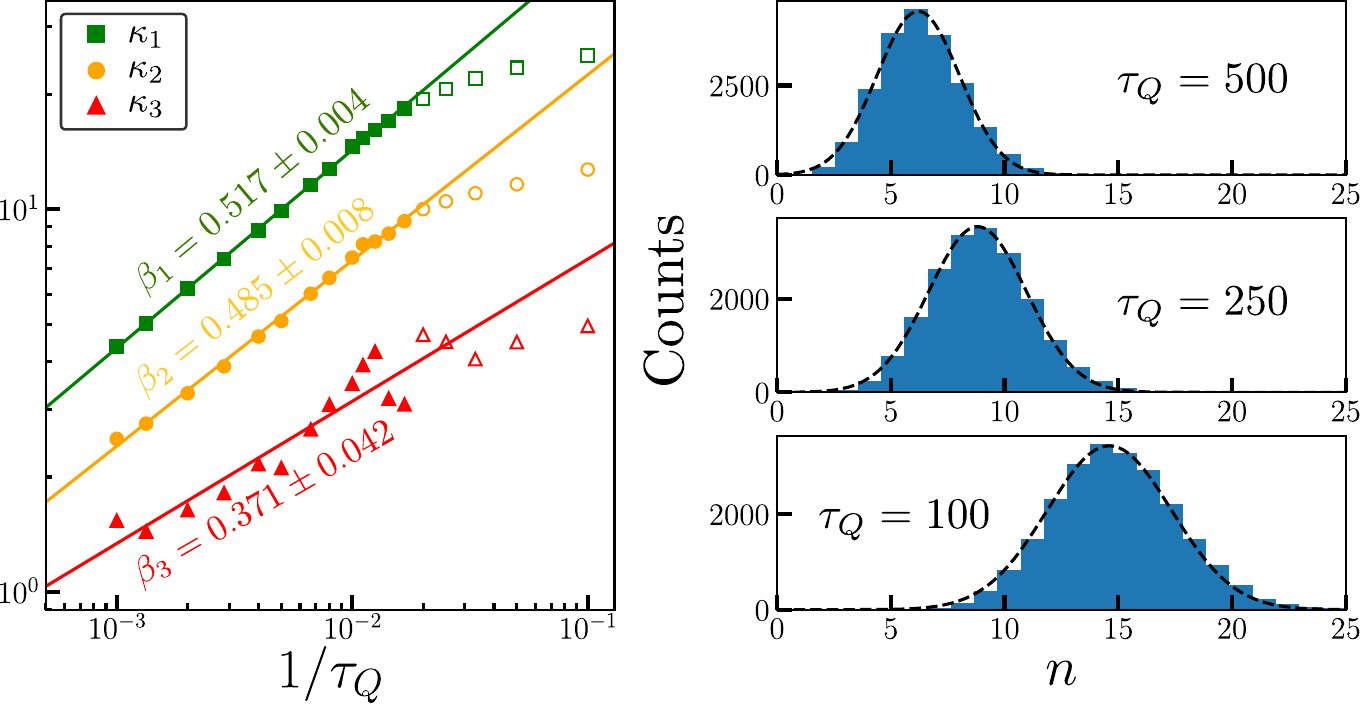}
\caption{Counting statistics and moments of the probability distribution for $2\pi$ phase vortices in component $A_x$. The left panel shows KZ scaling of the first three cumulants of the probability distribution for the number of phase vortices: mean number $(\kappa_1)$, variance $(\kappa_2)$, and skewness, $(\kappa_3)$, as a function of $1/\tau_Q$. Note the deviation of the scaling exponents from one another and the mean-field value, $\beta=1/2$. For each quench rate we carried out $N=20000$ quenches with random noise realizations, sufficient to capture the scaling of these cumulants. The dependence of scaling exponent $\beta_3$ on the number of random noise realizations, $N$, approaches a stable mean value for $N\ge 10000$, with fractional error of $\delta \beta_3 / \beta_3 \sim 10\%$ for $N=20000$. For slow quenches, $1/\tau_Q<10^{-3}\btauGL^{-1}$, full counting statistics were not computationally accessible. 
The right panel shows the number distribution of vortices for quench rates: $\btauGL/\tau_Q=500,250,100$, with Gaussian approximates (dashed lines). For faster quenches the mean number of defects increases and the distribution spreads. The quench rate (time) is in units of $1/\btauGL$ ($\btauGL$).
}
\label{fig:kibble_zurek}
\end{figure}

To identify and measure the number of phase vortices at freeze-out, the winding number of the phase of the order parameter is computed around plaquettes of 4 spatial grid points throughout the computational domain. We count the number of $\pm 2\pi$ phase windings of the component $A_x$. At freeze-out domain wall structures are not well developed or identifiable, c.f. App.~\ref{app-animation}.
In Fig.~\ref{fig:kibble_zurek} we show results for the dependence of the mean ($\kappa_1$), variance ($\kappa_2$) and skewness ($\kappa_3$) of the distribution of $2\pi$ phase vortices as a function of the quench rate $1/\tau_Q$, along with power law fits (solid lines) that demonstrate KZ scaling over an order of magnitude or more in quench rate. Deviations from KZ scaling appear for the most rapid quenches; the number of defects saturates for sufficiently fast quenches, $1/\tau_Q\gtrsim 10^{-2}\times\btauGL^{-1}$, when the mean number of defects is determined by the coherence length at the final temperature. 
The KZ scaling is found to be independent of pressure, encoded in the strong-coupling $b_p^{\text{sc}}$ parameters.

Histograms, each based on $20,000$ quenches, of the number of phase vortices are shown for quench times $\tau_Q/\btauGL=100,250,500$ with fits to Gaussian distributions. For each of the moments of the distribution, $\kappa_i$, the scaling exponents, $\beta_i$, deviate from the mean-field prediction of $\beta=1/2$.
We also find small, albeit measureable, deviations between the $\beta_i$, lending qualified support to the extended KZ full-counting statistics models.~\cite{del18,gom20}

\begin{figure*}
\centering
\includegraphics[width=0.9\linewidth]{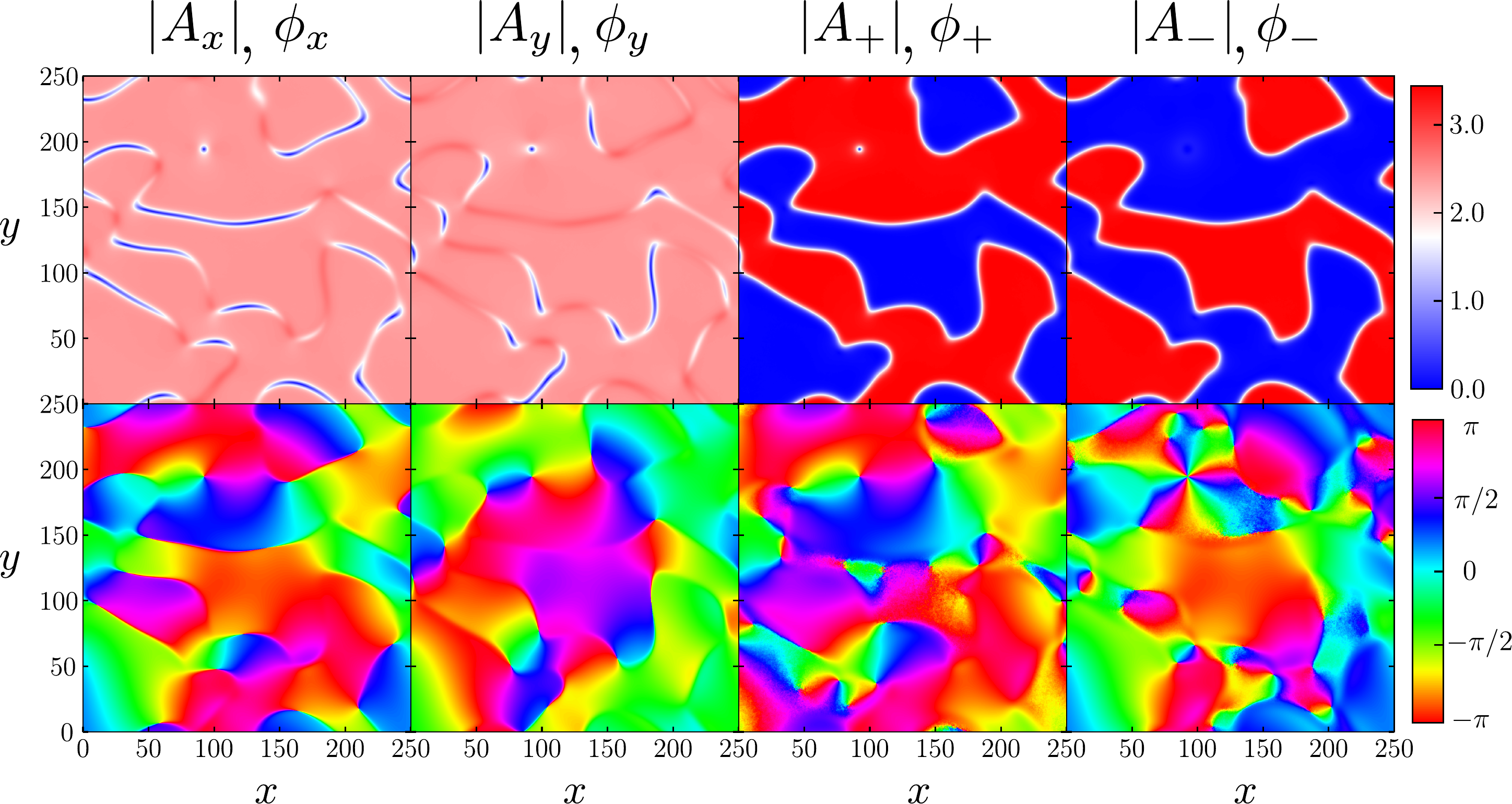}
\caption{Late-stage development of order order parameter for $t=399.5\;\btauGL$ at the temperature $T=0.5\,T_c$. The first two columns show the amplitude and phase of the Cartesian components of the order parameter. Amplitudes are in units of $k_B T_c$. The right two columns show the same order parameter resolved in the chiral basis $A_+$ and $A_-$. Domain wall and vortex structures are visible and described in the text. Coordinates $x$ and $y$ are in units of $\bxiGL$.
In order to show large-scale domain features, the computational domain was enlarged to $250 \bar{\xi}_{\text{GL}} \times 250 \bar{\xi}_{\text{GL}}$.
}
\label{fig:snapshot}
\end{figure*}

In the later stage evolution post-freeze-out, well defined domains of the two degenerate chiral phases emerge, separated by domain walls, with $\pm 2\pi$ phase vortices embedded in each domain. Vortices are also embedded on domain walls, leading to complex structure to domain walls (c.f. video animation in Fig.\ref{fig-animation} of App.~\ref{app-animation}). All of this structure is shown in Fig.~\ref{fig:snapshot} for a late-stage ($t=399.5\btauGL$) snapshot of the order parameter configuration.
In the first two colums we show the spatial map of the amplitude (1$^{st}$ row) and phase (2$^{nd}$ row) of the Cartesian components, $A_{x,y} =|A_{x,y}|e^{i\phi_{x,y}}$.

The largest scale defect structures are domain walls at the boundary of $A_+$ and $A_-$ chiral domains. The domain walls can be traced by regions of suppressed or enhanced amplitude in the Cartesian basis, and by suppressed amplitudes in the chiral basis. 
Chiral domains correspond to fixed relative phase, $\phi_y-\phi_x=\pm\pi/2$, which switches sign at a domain wall. As a result, chiral domains, domain walls and their evolution are most clearly tracked by monitoring the amplitude and phase of the chiral components, $A_\pm(\vr,t)$, as shown in the right two columns of Fig.~\ref{fig:snapshot}.
In a given chiral domain, e.g. $A_+$, the time-reversed order parameter, $A_-$,is suppressed, and as a result shows a noisy phase profile. 
At a domain wall between time-reversed chiral domains, the relative phase $\phi_y - \phi_x$ can change by either $\pm\pi$, such that there are two types of domain walls. At the interface of two distinct domain walls there is a $\pm 2\pi$ phase winding in either $A_x$ or $A_y$, related to vortices on the DW in the chiral basis (see App. \ref{DW-vortices}).
These embedded phase vortices are clearly shown in the plots of the phases $\phi_{x,y}$.
In this late stage very few phase vortices remain embedded in the regions of chiral domains. An isolated $+2\pi$ phase vortex is shown embedded at position $x=92.25\,\bxiGL\,,y=194.50\,\bxiGL$ in the $A_+$ chiral domain.
A novel feature of this vortex appears in the phase of the time-reversed order parameter, i.e. $\phi_-$, at the same position, i.e. a phase vortex with a winding number $m=+3$. This structure is a feature of phase vortices in a chiral ground state which we discuss in more detail in the next section.

\vspace*{-10mm}
\subsubsection{Structure \& Statistics of Vortices \& Anti-Vortices}
\vspace*{-3mm}

The nonequilibrium dynamics of the order parameter during and after a temperature quench that starts above $T_c$ then proceeds at a fixed quench rate through the transition to a final temperature below $T_c$ separates into three periods: an ``inertial period'' which is fluctuation dominated with an average order parameter that is nearly zero, followed by a narrow time window of a rapid growth of order parameter field which we identify with the time-frame of KZ freeze-out, then the post-freeze-out evolution dominated by the dynamics of topological defect structures - vortices and domain walls.
Video animation of the time evolution of the order parameter starting from thermal fluctuations of the order parameter at an initial temperature $T=1.5\,T_c$, followed by the dynamics during and after the quench, and finally the late-stage dynamics of topological defects, is included in the App.~\ref{app-animation}.

Vortex structures and precursor domains emerge soon after freeze-out. Vortices form in regions that evolve into chiral domains, but often vortices form on emerging domain walls.
The post freeze-out dynamics is complex, with domain walls capturing vortices, vortices and anti-vortices anihilating, and chiral domains undergoing expansion or collapse. Isolated vortices emerge from the collapse of a chiral domain with defects on the domain wall (c.f. App.~\ref{app-animation}).
There are distinct vortex structures that emerge in the chiral domains. These structures have been identified and discussed as possible vortex states in chiral superconductors.~\cite{tok90,sau94,sau09}

\begin{figure*}
\centering
\includegraphics[width=0.9\linewidth]{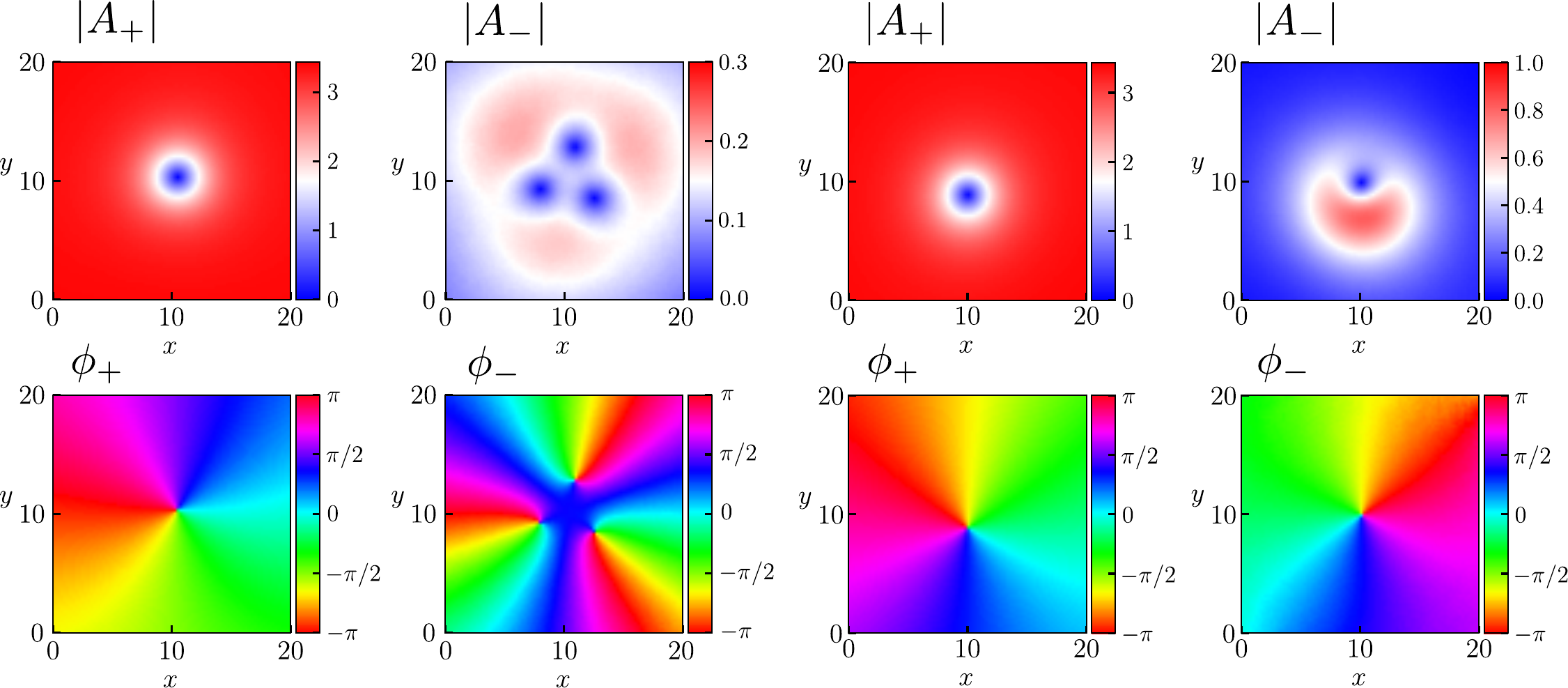}
\caption{
Structure of singly-quantized vortices with global winding numbers $p=\pm 1$ embedded in a chiral domain with $\vA\sim\ve_+$.
For $p=+1$ the amplitude $|A_+|$ has a topologically enforced node near the center of the cell, and the phase, $\phi_+$, winds counter-clockwise from $-\pi$ to $+\pi$ (column 1). The sub-dominant time-reversed order parameter has total winding number $m=+3$ represented by three satellite phase vortices, each with winding number $+1$ forming the triangular core structure (column 2).
For $p=-1$ the amplitude $|A_+|$ has a node near the center of the cell, and the phase, $\phi_+$, winds clockwise by $2\pi$ (column 3). However, the time-reversed order parameter, $A_-$, has winding number $m=+1$, and the singularity is off-set from that of $A_+$ leading to the crescent core structure (column 4).
Spatial coordinates are in units of $\bxiGL$, and the amplitudes $|A_\pm|$ are in units $k_B T_c$ at $T/T_c=1/2$.} 
\label{fig-vortex_structures_p1}
\end{figure*}

For an isolated vortex with global phase winding $p$, embedded in a chiral domain with $\vA\sim\ve_+$, the order parameter takes the form,
\begin{equation}
\vA(\vr) = 
\Delta\left[
a_+(\vr)
\,e^{i p \phi}\,\ve_+
+ 
a_-(\vr)
\,e^{i m \phi}\,\ve_-
\right]
\,,
\end{equation}
with $a_+(\vr)\sim r^{|p|}$ and $a_-(\vr)\sim r^{|m|}$ for $r\ll\bxiGL$ assuming axial symmetry is locally preserved. 
Far from the core we have $a_+(\vr)\rightarrow 1-\alpha_+\,r^{-2}$ and $a_-(\vr)\rightarrow \alpha_-\,r^{-s}$ with $s\ge 2$ for $r\gg\bxiGL$.
Most importantly, the phase winding of the time-reversed vortex core amplitude is constrained by axial symmetry in the far field: $p+1=m-1$.~\cite{sau94} This implies the \emph{inequivalence} of vortices and anti-vortices in a fixed chiral phase; i.e. a vortex with $p=+1$ in the $\ve_+$ phase has a core structure with winding number $m=+3$. However, the anti-vortex with $p=-1$ in the same $\ve_+$ phase has a core structure with winding number $m=+1$. These two vortex states are non-degenerate as the core energies of the two structures are necessarily different.
In fact the gradient and condensation energies favor the breaking of axial rotation symmetry for both singly-quantized vortices. 
As shown in the left two panels of Fig.~\ref{fig-vortex_structures_p1} for the $p=+1$ vortex the time-reversed core amplitude dissociates into \emph{three} singly quantized vortices bound to the core in a triangular structure. This structure is energetically more favorable than the axial $m=+3$ vortex as the former allows for the recovery of suppressed condensation energy.
Similar energetics leads to broken axial symmetry of the $p=-1$ vortex. In this case the node of the time-reversed amplitude, $A_-$, is displaced relative to the node of the bulk chiral amplitude, $A_+$, leading to the \emph{crescent} vortex structure shown in the right two panels of Fig.~\ref{fig-vortex_structures_p1}. The complete list of singly quantized vortices is given in Table \ref{vortex_table}.

In contrast to a single-component superfluid, doubly quantized ($p=\pm 2$) vortices in a two-fold degenerate chiral condensate are topologically stable. The winding numbers for the global phase and core amplitude $(p,m)$ are: $(+2,+4)$ and $(-2,0)$. 
Thus, dissociation of a $(+2,+4)$ vortex into a pair of $(+1,+3)$ vortices is topologically prevented within a bulk, single domain chiral phase.   
The anti-vortex with $p=-2$ is novel in that it allows the time-reversed phase to fill the vortex core without suppression since there is no phase winding for the core amplitude. This can lead to energetic stability of an array of doubly quantized vortices in a chiral ground state.~\cite{ich12}
However, in our simulations we have observed \emph{only} singly-quantized vortices and anti-vortices.

Another novel vortex structure is created near an impurity, or a mesoscopic structure, embedded within a chiral domain, e.g. an electron bubble. In this situation suppression of the bulk chiral order parameter near the impurity is accompanied by nucleation of the time-reversed chiral order parameter in order to generate a non-chiral order parameter at the surface of the impurity. The time-reversed order parameter carries a phase winding corresponding to a defect with winding numbers $(0,+2)$, i.e. the defect acquires a phase winding and thus supports an edge current.~\cite{sau09,she16}
The edge current also reflects the discontinuous change in topology from the 
bulk chiral phase with Chern number $N=+1$ to the topologically trivial ($N=0$) embedded mesoscopic impurity.
Similarly, domain walls separating time-reversed chiral phases with different Chern numbers, $N=+1$ and $N=-1$, also require one of the order parameter components to vanish on the domain wall, and as a result currents are expected to flow along the domain wall.~\cite{wu23}

\vspace*{-8mm}
\subsubsection{Mass Currents \& Topological Defects}
\vspace*{-3mm}

\begin{figure}
\centering
\includegraphics[width=0.95\linewidth]{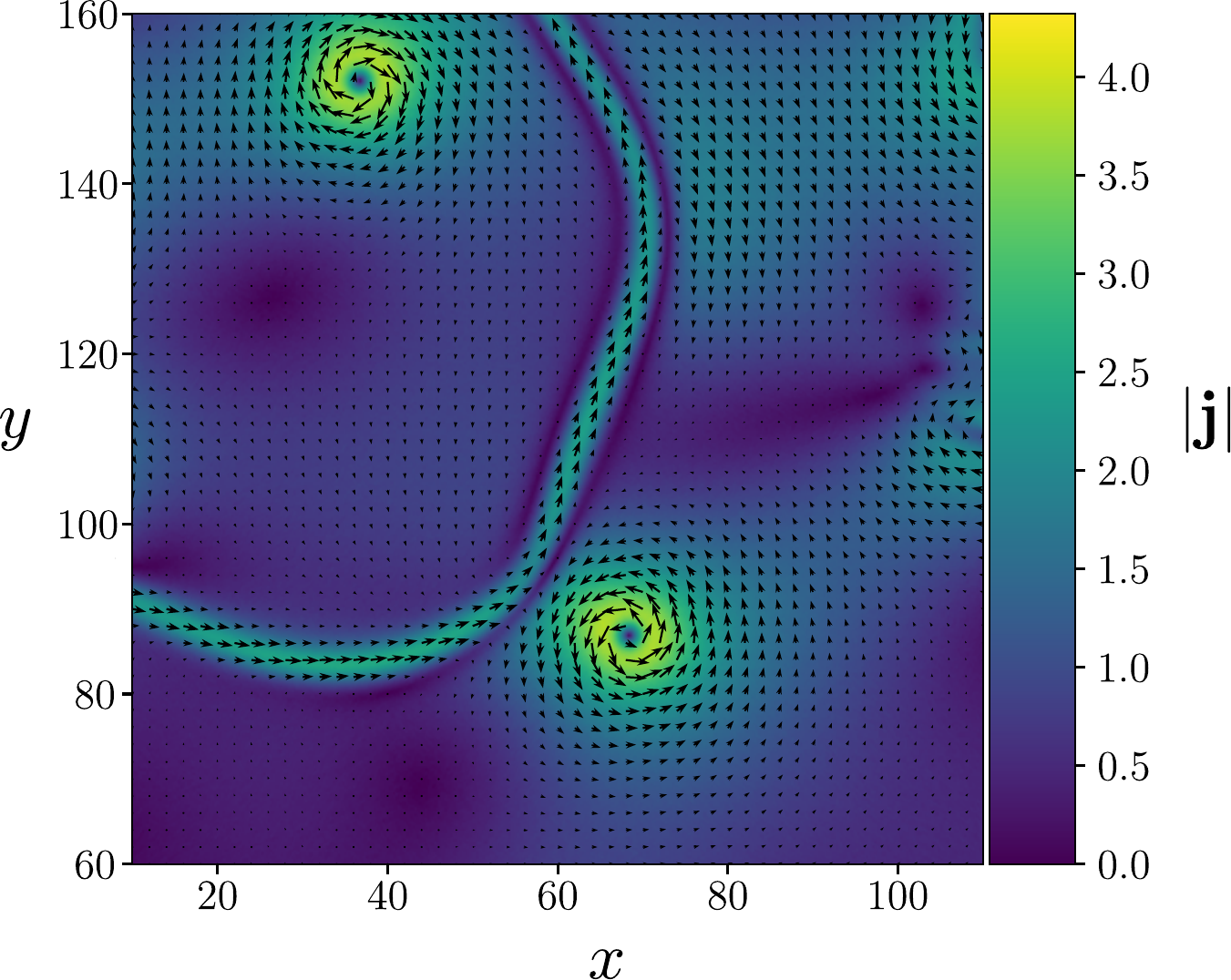}
\caption{Mass currents associated with vortices and domain walls between distinct chiral domains from late-stage evolution post freeze-out. The arrows denote the direction of the current, while the color bar indicates the magnitude of the current density.
The $p=+1$ vortex at position $(68.5,87.0)\bxiGL$ is embedded in a $\vA\sim\ve_+$ chiral domain, while the $p=-1$ vortex at $(36.75,152.0)\bxiGL$ is embedded in a time-reversed $\vA\sim\ve_-$ domain. The triangular core structures of these two vortices are unresolved in this image.
Spatial coordinates are in units of $\bxiGL$, and the current density is in units of $4m K_1 (k_B T_c)^2 / \hbar \bxiGL$.
}
\label{fig-Currents}
\end{figure}

For an inhomogeneous phase generated by a nonequilibrium quench transition, the phase generally hosts multiple chiral domains. The domain walls support topological edge currents which we can calculate from the mass current density obtained from TDGL Lagrangian,
\begin{equation}
j_i =
\frac{4m K_1}{\hbar}
\textrm{Im}
\left(A_j^*\partial_j A_i + A_j^* \partial_i A_j + A_i^*\partial_j A_j\right)
\,,
\end{equation}
where $m$ is the mass of the \He\ atom.
Fig.~\ref{fig-Currents} shows the circulating currents of a vortex and an anti-vortex in two distinct chiral domains which are time-reversed partners of one another. The chiral edge currents confined on the domain wall separating the time-reversed chiral domains are also clearly visible.

\begin{table}
\begin{center}
\begin{tabular}{|c c c c c|} 
 \hline
 $A_+$ & $A_-$ & $p$ & $m$ & Label \\ [0.5ex] 
 \hline\hline
 Dominant & Subdominant & $+1$ & $+3$ & Triangle \\ 
 \hline
 Dominant & Subdominant & $-1$ & $+1$ & Crescent \\
 \hline
 Subdominant & Dominant & $-3$ & $-1$ & Triangle \\
 \hline
 Subdominant & Dominant & $-1$ & $+1$ & Crescent \\
 \hline
\end{tabular}
\caption{Types of bulk vortices and associated winding numbers. The ``Dominant'' label is for the bulk chiral domain the vortex is embedded in, and the ``Subdominant'' component only obtains nonzero amplitude near the core. The labels $p$ and $m$ refer to the winding numbers of $A_+$ and $A_-$, respectively. Vortex structures for the first two rows are shown in Fig. \ref{fig-vortex_structures_p1}, and the remaining two structures are time-reversed partners of the first two.}
\label{vortex_table}
\end{center}
\end{table}

\vspace*{-5mm}
\subsubsection{Statistics of Vortices \& Anti-Vortices}
\vspace*{-3mm}

The inequivalence of the core structure and excitation energy of vortices and anti-vortices in a chiral domain is distinct from that of scalar $\point{U(1)}{\ns}$ theory. In the $\point{U(1)}{\ns}$ case one expects - and finds - the same KZ scaling relation and probability distribution for the generation of vortices as that for anti-vortices following a thermal quench.
Does this symmetry between vortex and anti-vortex distributions hold for a nonequilibrium transition to a chiral phase?

To address this question we need a large enough computational domain to generate a sizeable population of vortices for each trial quench. For these calculations we carried out dynamical simulations on a $1000\,\bxiGL\times 1000\,\bxiGL$ domain.
Vortex and anti-vortex populations are counted after freeze-out when chiral domains are sufficiently well defined. We averaged the vortex populations over 100 quenches, i.e. different random noise realizations.

Identification of vortex cores is based on detecting localized regions of strongly suppressed order parameter amplitude using the \emph{Difference of Gaussian} (DoG) image processing algorithm.~\cite{low04} Once these sites are located, the identity of the chiral domain is then found by measuring the amplitudes, $|A_\pm|$, for a large region surrounding the vortex core. Finally, the phase winding is computed on a small plaquette surrounding the core. This three-step process provides a unique identification of triangular and crescent vortices for either chiral domain, which are counted via the classification in Table \ref{vortex_table}. This approach with image processing is efficient for finding bulk vortex cores that are near other defects (domain walls or other vortex cores) while excluding domain wall defects.

\begin{figure}
\centering
\includegraphics[width=0.98\linewidth]{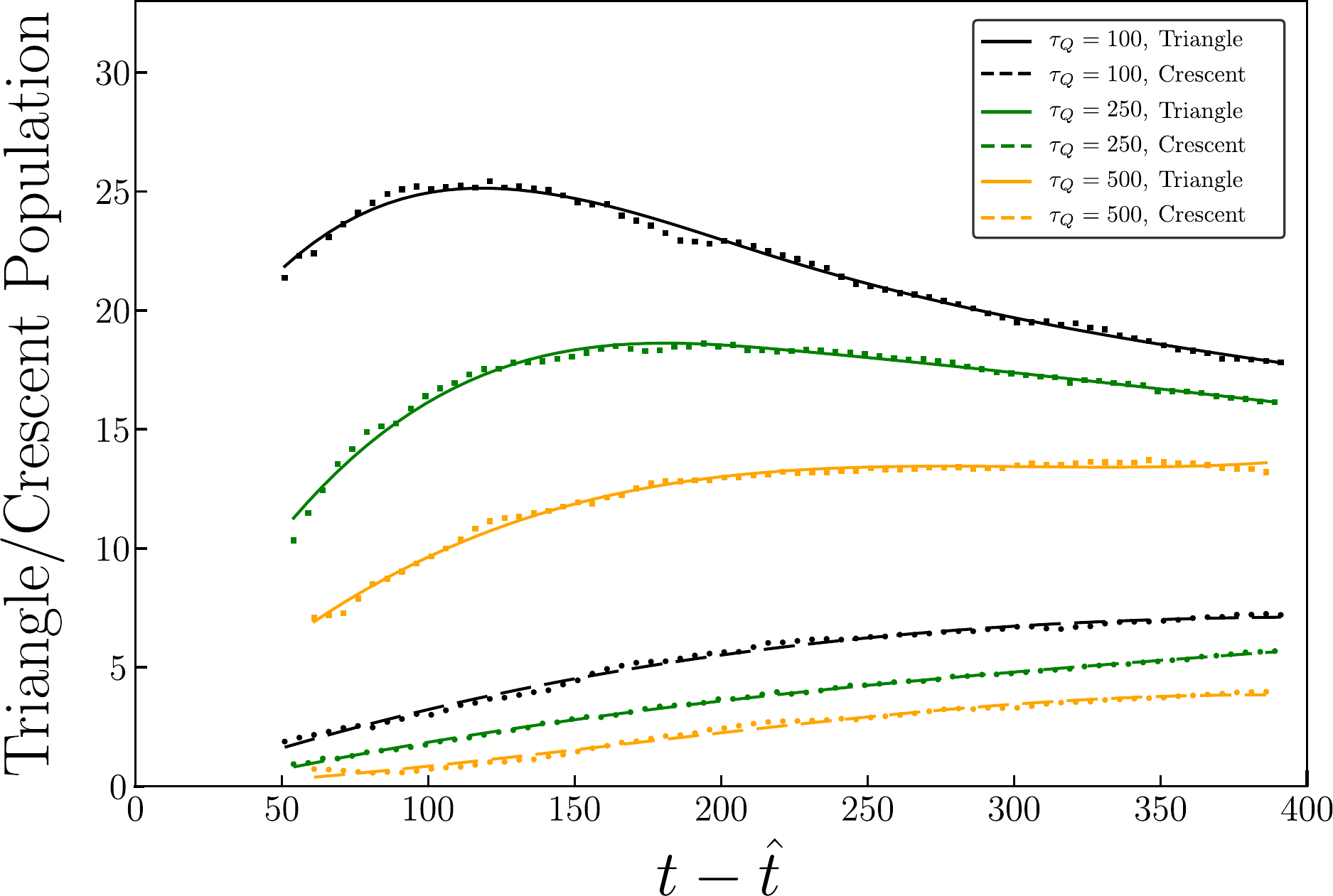}
\caption{Asymmetry in the number of singly-quantized vortex structures for the inhomogeneous chiral phase of $^3$He post freeze-out at $p=5.0$ bar. Results based on dynamical simulations are shown for a $1000\,\bxiGL\times 1000\,\bxiGL$ computational domain averaged over 100 quenches. Solid and dashed lines are spline fits to triangle and crescent vortex data, respectively. Time is in units $\bar{\tau}_{\text{GL}}$.
}
\label{fig-vortex_asymmetry}
\end{figure}

Fig.~\ref{fig-vortex_asymmetry} shows the results for triangular (solid splines) and anti-vortex (dashed splines) populations as a function of time after freeze-out, $t-\hat{t}$, for quench times $\tau_Q/\btauGL = 100,\,250,\,500$. There is a significant asymmetry favoring triangular vortex structures over crescent vortex structures. Note that both vortex structures are present in both chiral domains as time-reversed partners, i.e. the time-reversed partner to the triangular $p=+1$ vortex in a $\vA\sim\ve_+$ chiral domain, is the $p=-1$ anti-vortex in a $\vA\sim\ve_-$ chiral domain, and similarly for crescent vortices.
Thus, statistically the number of vortices and anti-vortices will be on average equal provided the mean areal densities of the two chiral domains are equal.

However, the question remains as to \emph{why is there a significant asymmetry in the number of triangular vortices compared to crescent vortices?}
To put this question in context first consider times just prior to freeze-out in which according to the KZM patches of order parameter fluctuations with random phases are causally disconnected. At freeze-out if there is net a winding number connecting a group of patches, a precursor vortex can form, however, at this early stage of vortex formation there is no time for the formation of a core structure nor a background chiral domain, At freeze-out there is no bias in the sign of the phase winding connecting patches of order parameter, nor an \emph{a priori} bias for vortices to evolve into triangular or crescent structures.

Fig.~\ref{fig-vortex_asymmetry} shows the asymmetry in the number of triangular versus crescent vortices, which appears to grow rapidly from freeze-out. While we are unable to differentiate these two classes of vortices for early post-freeze-out times, we do observe a significant increase in the number of the triangular vortices for $t-\hat{t}\gtrsim 50\,\btauGL$. One possibility suggested by the numerical data is that the asymmetry onsets rapidly post freeze-out from an asymmetry in the interaction of the two classes of vortices with domain walls, and/or the preferrential capture of crescent vortices. This conjecture needs further study using larger scale simulations and refined visualization algorithms.

\vspace*{-5mm}
\section{Summary and Outlook}
\vspace*{-3mm}

\begin{figure*}
\centering
{
\href{https://arxiv.org/src/2412.03544v1/anc/Quench_3He-Film.mp4}
{\includegraphics[width=\textwidth]{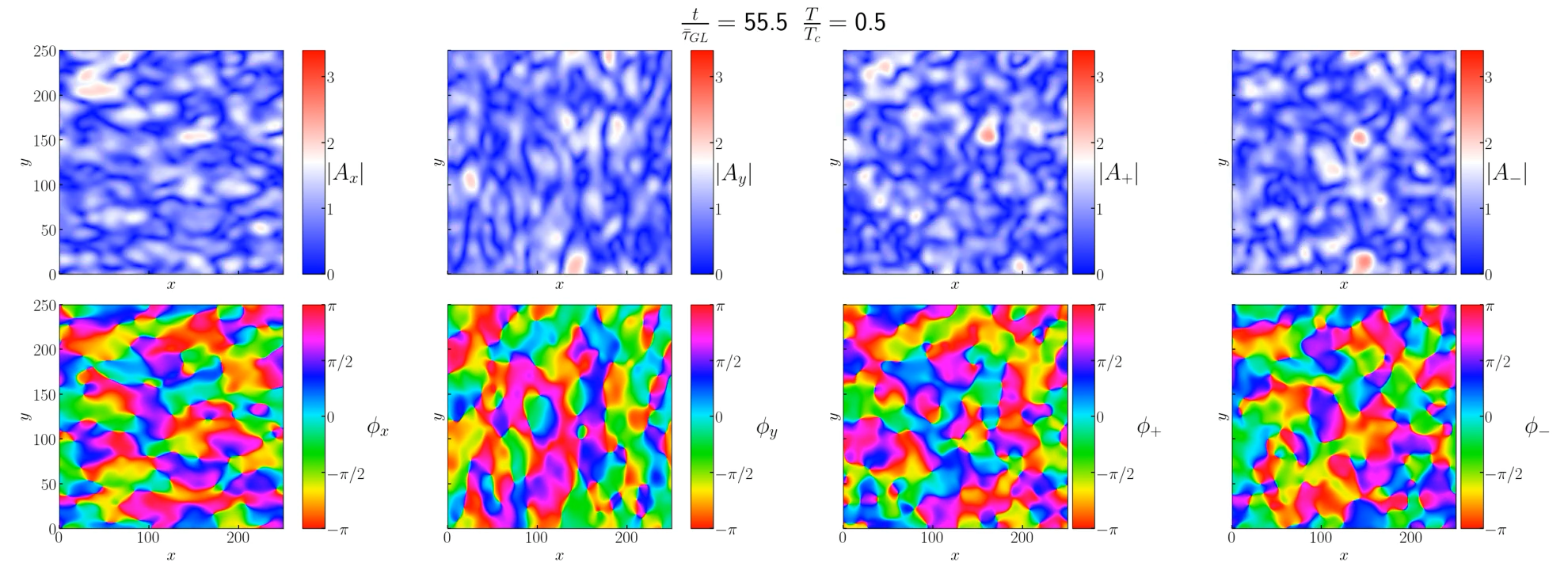}}
}
\caption{
An early time frame of an animation of 
order parameter dynamics for a thin film of \Hea\
undergoing a temperature quench with $\tau_Q/\btauGL=100$,
and evolving under the dynamics of the TDGL equations
coupled to thermal bath described by Langevin noise and dissipation.
%
Click on the image to view or download the mp4 video using your default browser. 
Key timeframes (in $\btauGL$) are:
(a) $t=0$ - passage through $T_c$,  
(b) $t=47.5\,\btauGL$ - end of ``inertial period'' with small amplitude, 
(c) $t=50.0$ - final temperature of the quench at $0.5\,T_c$,
(d) $t=60.0$ - KZ freeze-out time,
(e) $t=63.5$ - well defined chiral domains form, 
(f) $t=119.5$ - coarsening of domains and phase structures,
(g) $t=107.5-151.5$ - collapse of an $A_-$ chiral domain,
(h) $t=163.5-303.5$ - shrinkage of an $A_-$ chiral domain to a $2\pi$ vortex,
(i) $t=235.5-303.5$ - annihilation of a vortex in $A_-$ on a DW.
\label{fig-animation}
}
\end{figure*}

We carried out studies of order parameter dynamics and the generation of topological defects in thin films of superfluid \He\ undergoing a temperature quench through the superfluid transition. Our results are based on a time-dependent generalization of Ginzburg-Landau theory for strong-coupling \He\ that includes the dissipative dynamics resulting from the coupling of the order parameter to thermal noise. 

The ground state for thin \He\ films is the A-phase, which is doubly degenerate with the chiral axis parallel or anti-parallel to the normal to the film surface, i.e. $\hat{\ns\vell}||\pm \hat\vz$. As a result superfluid \Hea\ films host both quantized vortices and domain walls separating degenerate chiral domains.
Numerical simulations of the nonequilibrium phase transition support the KZM for topological defect generation. The freeze-out time and mean number of vortices obey the KZ scaling relations, as do higher order moments of the vortex number distribution.
Our results for full counting statistics of vortices show small, but measurable, differences in the scaling exponents for the first three moments of the vortex number distribution. 
Simulations of post-freeze-out dynamics show interactions between vortices and domain walls; vortices bound to domain walls, vortex/anti-vortex annihilation on domain walls, collapse and growth of domain walls, including shrinking of domain walls to vortices.
Vortices and anti-vortices in a chiral domain are described by a pair of winding numbers, $(p,m)$, for the phase of the far field amplitude ($p$) and the near-field (core) amplitude ($m$) that are related by $p+l=m-l$ where $l=\pm 1$ is the Chern number the chiral ground state. Thus, vortices and anti-vortices are inequivalent, differentiated by the structure of their vortex cores, either ``triangular'' or ``crescent''.
Our large-scale quench simulations show a large asymmetry in the number of triangular versus crescent vortices post-freeze-out which we argue reflects complex interactions and dynamics between vortices and domain-walls.
Lastly, the simulations we report are based on a tested strong-coupling GL functional. Thus, we hope our predictions for topological defect generation, KZ scaling and statistics can be tested experimentally under conditions of fast cooling of \He\ confined in slabs or thin films.

\noindent\underline{\sl Acknowledgements:} 
We thank Smitha Vishveshwara for discussions on Kibble-Zurek physics in chiral superconductors that in part motivated this study. We thank Mark Hindmarsh and Quang Zhang for discussions on nonequilibrium phase transitions in superfluids and quantum field theory.
This research was supported by the Hearne Institute of Theoretical Physics and the Center for Computation and Technology at Louisiana State University.

\vspace*{-8mm}
\appendix
\section{Temperature Quench for a \Hea\ Film}\label{app-animation}
\vspace*{-5mm}

Figure~\ref{fig-animation} shows an early time frame of the simulation of a thermal quench. The mp4 video animation of the dynamics is linked to the image and here: \href{https://arxiv.org/src/2412.03544v1/anc/Quench_3He-Film.mp4}{quench animation}.

The temperature quench starts $T=1.5\,T_c$ (time $t=-50\,\btauGL$) crosses $T_c$ at $t=0$
cooling at a rate $\tau_Q^{-1} = 10^{-2}\btauGL^{-1}$ down to a final temperature of $T=0.5\,T_c$ at $t=50\btauGL$.
After the temperature drops below $T_c$ increased structure in the phases of the order parameter develops with the amplitude remaining very small during an ``inertial period'' from $t=0$ to 
$t=47.5\,\btauGL$.
Growth of the amplitude starts near the end of the quench at $t=50.0\,\btauGL$, followed by rapid growth which we identify with KZ freeze-out near $t=60.0\,\btauGL$. At 
$t=63.5\,\btauGL$ the chiral amplitudes, $|A_+|$ and $|A_-|$ show the early development of chiral domains.

Later stage evolution at $t=119.5\,\btauGL$ shows coarsening with well defined chiral domains. The domain walls (DWs) show regions of strong curvature. The internal structure of a DW changes along the wall; phase vortices of the $A_x$ and $A_y$ components are trapped on the DW as shown in both the amplude and phase plots of the Cartesian amplitudes.

The dynamics of chiral domains, DWs and phase vortices is rich. For example, consider the two $A_-$ chiral domains at positions $(50,220)\bxiGL$ (small domain) and $(100,200)\bxiGL$ (large domain), and watch their evolution. The small domain collapses and vanishes, while the larger domain collapses and forms a $2\pi$ phase vortex embedded in a large $A_+$ chiral domain, visible in both amplitude and phase at $t=267.5\,\btauGL$. The $A_-$ core of this embedded vortex has the $6\pi$ phase winding of the triangular vortex structure.

Note the embedded vortex in the $A_-$ domain near a DW at position $(140,140)\bxiGL$. This vortex is attracted to the DW, annilates with a vortex bound to the DW at $t=303.5\,\btauGL$ and in the process alleviates the sharp curvature of the DW at that location.

\eject

\noindent The final frame at $t=399.5\,\btauGL$ is shown in Fig.~\ref{fig:snapshot}, and described in the text and caption.

\vspace*{-5mm}
\section{Structure of Defects on DW}\label{DW-vortices}
\vspace*{-5mm}

\begin{figure*}
\centering
\includegraphics[width=0.95\linewidth]{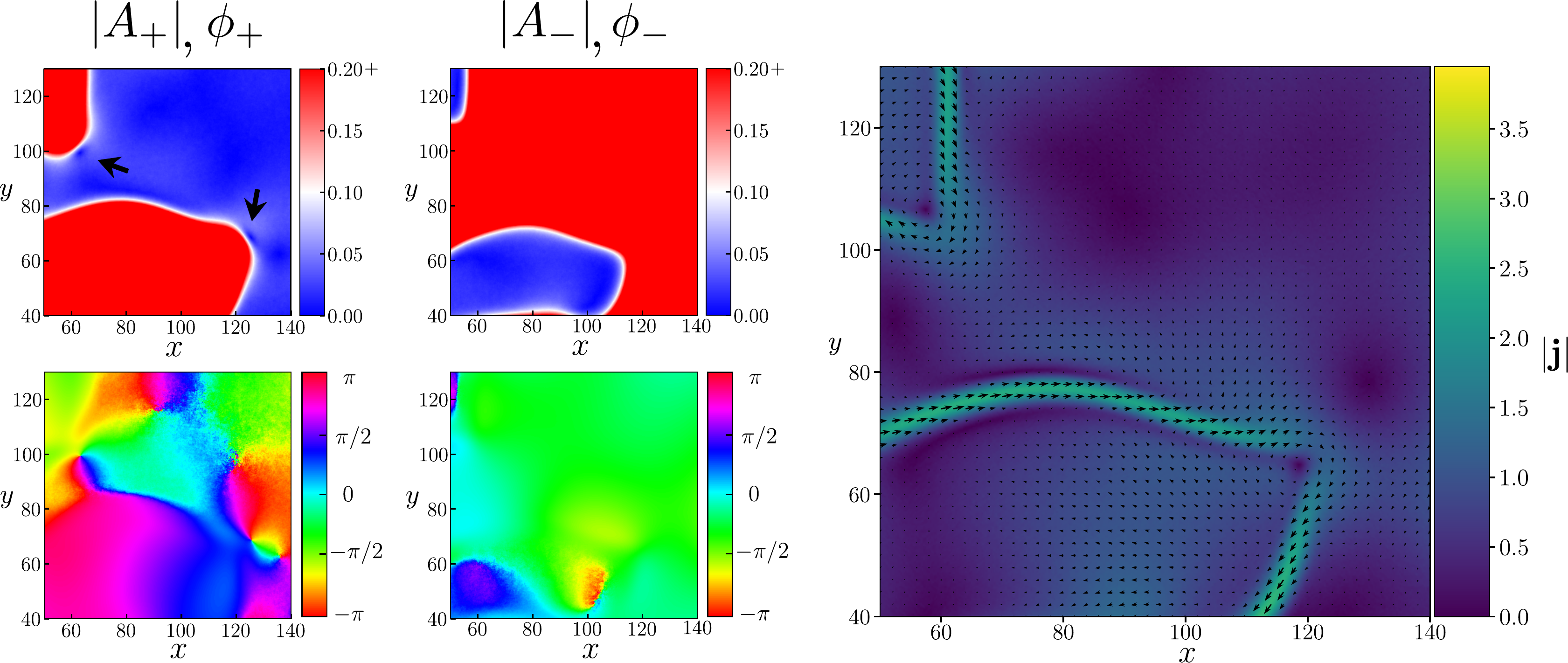}
\caption{Structure of topological defects trapped by a domain wall. 
Left panel: the four plots are the amplitude and phase of order parameter components $A_+$ and $A_-$. 
In the amplitude plots, The color scale for the amplitude plots has been modified for clarity: solid red corresponds to amplitudes $|A_\pm| \geq 0.20\;k_B T_c$. 
In the $A_-$ region there are two vortices on the domain wall shown in the $A_+$ component at locations indicated by the black arrows in the upper left plot. The $A_+$ defects on the domain wall have $-2\pi$ phase windings. 
Right panel: the mass current density in the vicinity of the two topological defects. The DW edge currents are disturbed near near DW vortices. Current flows around a sharp corner, and there is suppressed current density near at the vortex core. The spatial scale is shown in units of $\bxiGL$, and the mass current density is shown in units of $4m K_1(\kb T_c)^2/\hbar\bxiGL$.
}
\label{edge_defects}
\end{figure*}

The structure of vortices trapped on the domain walls between chiral domains, 
hereafter ``DW vortices'' are distinct from vortices embedded in either chiral domain.
the structure of two DW vortices in the $A_+$ order parameter component is shown in Fig.~\ref{edge_defects}, as well as the influence of the DW vortices on the DW edge current. 
The DW vortices in the $A_+$ component exhibit $-2\pi$ phase windings 
and are pinned at the edge of the $A_-$ domain as shown by the black arrows
in Fig.~\ref{edge_defects}.
DW vortices in the $A_-$ component, while not shown in Fig.~\ref{edge_defects}, are time-reversed partners of the $A_+$ DW vortices, i.e. $+2\pi$ phase windings of the $A_-$ amplitude pinned just at the edge of $A_+$ domain.

DW vortices have amplitude profiles with two general features: (i) they exist on corners, or sharp edges, of the DW wall and (ii) they modify the DW edge current, such that there is continuity of current flow around the corner of the DW, and there is a suppressed current density at the core of the DW vortex.

\newpage

\vspace*{-5mm}
%
\end{document}